\documentclass{jps-cp}
\usepackage{xcolor}
\usepackage{color}

\newcommand{\tH}{t_{\scalebox{0.7}{h}}}
\newcommand{\etaL}{\eta_{\scalebox{0.5}{L}}}

\title{Photoinduced $\eta$-pairing in One-dimensional Mott Insulators}

\author{
Satoshi~\textsc{Ejima}$^{1,2}$, 
Tatsuya~\textsc{Kaneko}$^3$, 
Florian~\textsc{Lange}$^1$,
Seiji~\textsc{Yunoki}$^{2,4,5}$,
Holger~\textsc{Fehske}$^1$
}

\inst{
$^{1}$ Institut f\"ur Physik, Universit\"at Greifswald, D-17489
       Greifswald, Germany\\
$^{2}$ Computational Condensed Matter Physics Laboratory, RIKEN
       Cluster for Pioneering Research (CPR), Wako, Saitama 351-0198, Japan\\
$^{3}$ Department of Physics, Columbia University, New York, NY 10027, USA\\
$^{4}$ Computational Quantum Matter Research Team, RIKEN Center 
       for Emergent Matter Science (CEMS), Wako, Saitama 351-0198, Japan\\
$^{5}$ Computational Materials Science Research Team, RIKEN Center 
       for Computational Science (R-CCS), Kobe, Hyogo 650-0047, Japan
}

\email{ejima@physik.uni-greifswald.de}

\recdate{September 15, 2019}

\abst{
Employing the density-matrix renormalization group technique
in the matrix-product-state representation, we investigate the 
photoexcited superconducting correlations induced 
by the $\eta$-pairing mechanism in the half-filled Hubbard chain. 
We estimate the characteristic pair correlation function and
verify the accuracy of our numerical results by 
comparison with exact-diagonalization data for small systems.
The optimal parameter set of the pump that most enhances the $\eta$-pair  
correlations, is calculated in the strong-coupling 
regime. For such a pump, we explore the possibility of 
quasi-long-range order.   
}

\kword{superconducting correlations, Hubbard model, DMRG}

\begin{document}
\maketitle

\section{Introduction}

Superconductivity is one of the most fascinating phenomena 
in condensed-matter physics and has attracted continuous attention 
since its discovery in 1911~\cite{RevModPhys.66.763,RevModPhys.78.17}. 
While the basic features of the superconducting (SC) phase are 
well established for equilibrium systems, 
the photoinduced superconductivity is subject of intense research
because of the recent pump-probe experiments, e.g., in
copper oxides~\cite{Fausti189,Hu_2014,Nature_516_71} or 
in K$_3$C$_{60}$~\cite{Nature_530_430}. In these materials,
the transient optical spectra imply the SC-like properties even 
at elevated temperatures.

Very recently, it was theoretically demonstrated that 
unconventional SC correlations can be induced by 
pulse irradiation in a simple Mott insulator described 
by the half-filled Hubbard model~\cite{PhysRevLett.122.077002}. 
This SC state stems from the so-called $\eta$-pairing 
mechanism~\cite{PhysRevLett.63.2144,PhysRevLett.67.3848,ELER1992559}, 
characterized by staggered pair-density-wave oscillations 
in the off-diagonal long-range correlations~\cite{Fujiuchi-paper}. 
However, since these results were obtained by the exact diagonalization 
(ED) technique, the accessible system size was only 
$L\lesssim16$ with periodic boundary conditions 
(PBC)~\cite{PhysRevLett.122.077002}. 
It is, thus, vitally important to discuss the finite-size effects 
in the $\eta$-pairing state and pair correlations especially 
at larger distances.

To make progress in this direction, we focus on one-dimensional systems, 
where the unbiased density-matrix renormalization group (DMRG)
technique~\cite{White92}  can be applied. 
(We note that $\eta$-pairing state in the photoexcited state can also be found
in two dimensions~\cite{PhysRevLett.122.077002}, but is hard to calculate 
in an approximation-free way).
Simulating the real-space pair correlation function and its structure factor, 
we explore the conditions under which $\eta$-pairing is most likely. 
For large enough system sizes, we demonstrate a peculiar increase 
of the pair structure factor with the parameter set of maximally 
enhanced $\eta$-pairing state after the photo excitation, 
which might be the signature of the (quasi-)long-range order of pairs.

\section{Theoretical approach}
{\bf Model.} Hereinafter, we study the photoinduced $\eta$-pairing state 
in a half-filled Hubbard chain, defined by
\begin{eqnarray}
 \hat{H}=-\tH \sum_{j,\sigma} 
  \left(
   \hat{c}_{j,\sigma}^\dagger \hat{c}_{j+1,\sigma}^{\phantom{\dagger}}
   +\text{h.c.}
  \right)
  +U\sum_{j}\hat{n}_{j,\uparrow}\hat{n}_{j,\downarrow}\,,
\end{eqnarray}
where $\hat{c}_{j,\sigma}^{\dagger}$ ($\hat{c}_{j,\sigma}^{\phantom{\dagger}}$)
creates (annihilates) an electron with spin projection 
$\sigma=\uparrow,\downarrow$ 
at Wannier site $j$, and 
$\hat{n}_{j,\sigma}=\hat{c}_{j,\sigma}^{\dagger}
                    \hat{c}_{j,\sigma}^{\phantom{\dagger}}$.
The transfer amplitude $\tH$ enables the electrons to hop between 
neighboring lattice sites, whereas the on-site Coulomb repulsion $U$ 
tends to localize the electrons, establishing a Mott insulating ground state. 

Introducing a time-dependent external field into the hopping term 
by the Peierls substitution~\cite{Peierls1933},  
$\tH \hat{c}_{j,\sigma}^\dagger \hat{c}_{j+1,\sigma}^{\phantom{\dagger}} 
 \rightarrow 
\tH e^{\mathrm{i}A(t)} \hat{c}_{j,\sigma}^\dagger 
                       \hat{c}_{j+1,\sigma}^{\phantom{\dagger}} $, 
where $A(t)$ is the time-dependent vector potential  
\begin{eqnarray}
 A(t)=A_0 e^{-(t-t_0)^2/(2\sigma_{\rm{p}}^2)}\cos\left[\omega_{\rm{p}}(t-t_0)\right]
\end{eqnarray}
with amplitude $A_0$, frequency  $\omega_{\rm p}$ and
pulse width  $\sigma_{\rm p}$ centered at time $t_0$ ($>0$), 
makes the Hamiltonian time-dependent: 
$\hat{H}\to\hat{H}(t)$. As a consequence the equilibrium ground state
$|\psi(0)\rangle$ at $t=0$ evolves in time 
$|\psi(t)\rangle$. To take account of this time evolution, in our numerics, 
we apply the time-evolving block decimation 
method~\cite{PhysRevLett.91.147902} in combination with the second-order Suzuki--Trotter decomposition. 
In the following, we use $\tH$ ($\tH^{-1}$) as the unit of energy (time), 
and the time step $\delta t$ is set to be $\delta t\cdot \tH=0.01$.

{\bf Pairing correlations.} The photoinduced $\eta$-pairing state can be characterized 
by the real-space pair correlation function
\begin{eqnarray}
 P(r,t)=\frac{1}{N_{\rm b}}\sum_{j=1}^{N_{\rm b}}
  \langle\psi(t)|
   \left(
    \hat{\Delta}_{j+r}^\dagger \hat{\Delta}_{j}^{\phantom{\dagger}}
    +\rm{h.c.}
   \right)
   |\psi(t)\rangle
 \label{Eq-Prt}
\end{eqnarray}
with the on-site singlet pair operator
$\hat{\Delta}_j=\hat{c}_{j,\uparrow}\hat{c}_{j,\downarrow}$. 
Here, $N_{\rm b}=L-r$ denotes the number of pairs of sites separated by distance $r$
in a system of $L$ sites with open boundary conditions (OBC).  
For $r=0$ the pair correlation is equal to twice
the double occupancy, i.e., $P(r=0, t)=2n_{\rm d}(t)$, 
where $n_{\rm d}(t)=(1/L)\sum_{j=1}^{L}
  \langle\psi(t)|
   \hat{n}_{j,\uparrow} \hat{n}_{j,\downarrow}
  |\psi(t)\rangle$
as demonstrated in Ref.~\cite{PhysRevLett.122.077002}.
Note that it is important to analyze the 
Fourier transform of $P(r, t)$, i.e., 
$P(q,t)=\sum_r e^{\mathrm{i}q r}P(r,t)$, which shows a characteristic 
enhancement after the pulse irradiation in the half-filled Hubbard model. 

\begin{figure}[htb]
 \begin{minipage}[t]{.46\textwidth}
  \centering
  \includegraphics[width=\textwidth]{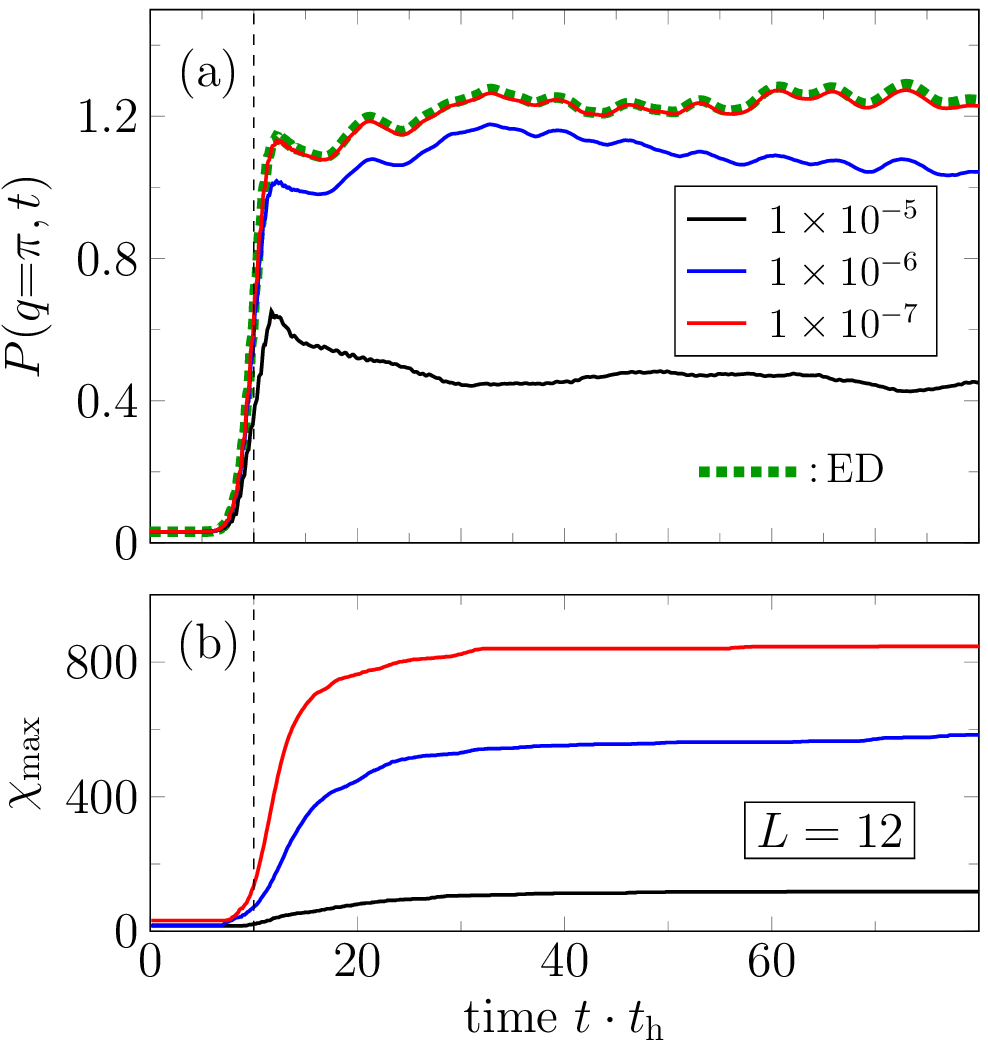}
 \end{minipage}
 \hspace{0.02\linewidth}
 \begin{minipage}[t]{.46\textwidth}
  \centering
  \includegraphics[width=\textwidth]{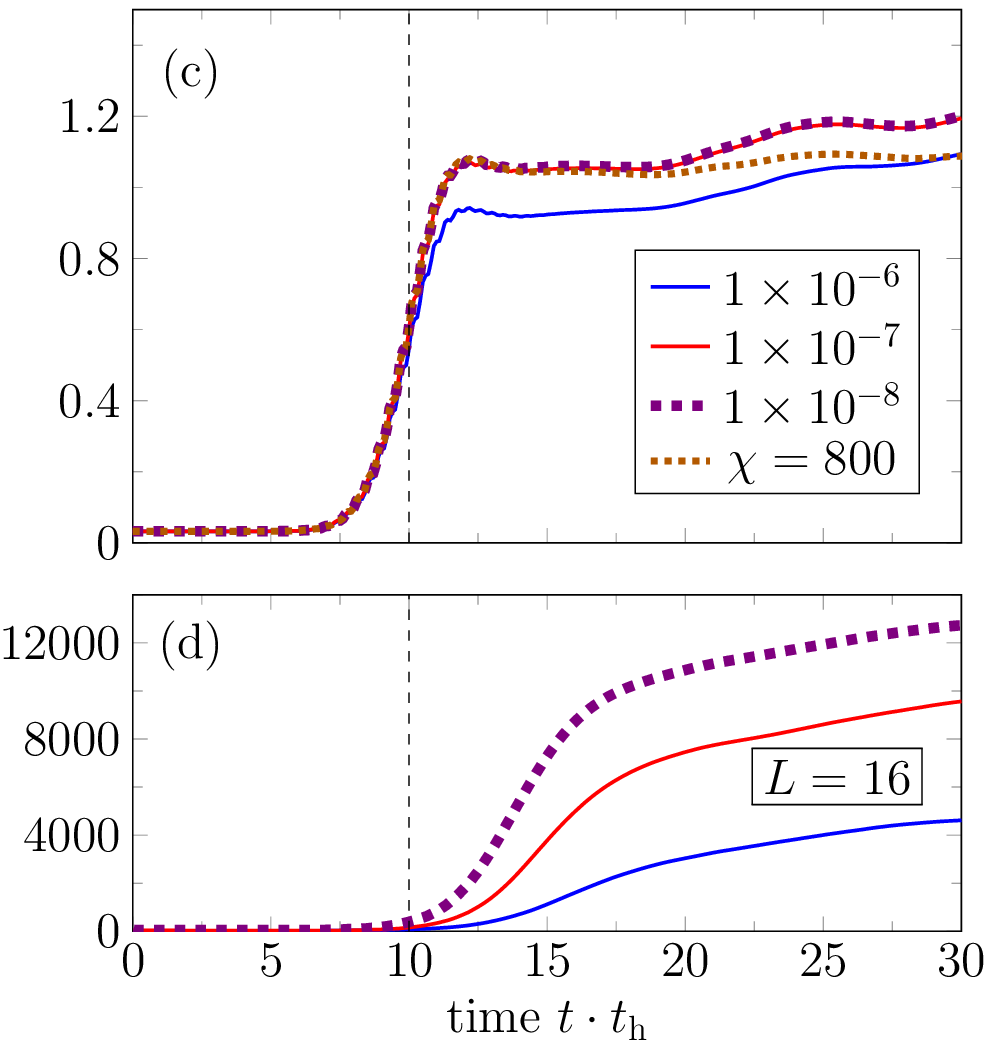}
 \end{minipage}
 \caption{
 Cutoff dependence of $P(q=\pi, t)$ for $U/\tH=8$ 
 within OBC DMRG (upper panels).  Corresponding maximum bond dimension $\chi_{\rm max}$ 
 needed to keep the maximum truncation error smaller than 
 some fixed values at any time (lower panels). 
 The dotted line in (a) is obtained by ED.
 $A(t)$ is parametrized by $A_0=0.4$, 
 $\omega_{\rm p}/\tH=8$, $\sigma_{\rm p}\cdot\tH=2$ and $t_0\cdot\tH=10$.}
 \label{Ppi-t}
\end{figure}

{\bf Accuracy check.}  Before we show our main results, 
we like to discuss the accuracy of the time-dependent DMRG simulations 
with OBC. During a DMRG simulation one chooses the maximum of  
the so-called bond dimension $\chi$ and makes sure that the truncation 
error in the singular value decomposition stays sufficiently small.
The other way around, the maximum truncation error can also 
be fixed in the simulation. 
Figure~\ref{Ppi-t} demonstrates the latter case, simulating 
the pair structure factor $P(q=\pi, t)$ by DMRG.  
The DMRG data are compared with ED results in Fig.~\ref{Ppi-t}(a)  for $L=12$. 
Keeping the cutoff smaller than $1\times 10^{-7}$, $P(q=\pi,t)$ 
obtained by DMRG is in perfect agreement with the ED results, even up to time 
$t\cdot\tH=80$. The deviation only becomes significant when a larger cutoff 
is used for $t\gtrsim t_0$. In each case, the maximum bond dimension
$\chi_{\rm max}$ increases rapidly around $t\sim t_0$ and afterwards stays constant
for $t\cdot\tH\gtrsim 25$ [see Fig.~\ref{Ppi-t}(b)].
With increasing system size, the memory required by ED calculations 
increases exponentially: It is already problematic
to carry out a time-dependent ED simulation for $L=16$. 
Clearly the computational costs of the time-dependent DMRG calculations also increases 
with the system size $L$, but the truncation error can be kept
below $1\times 10^{-8}$ in the case of $L=16$. To keep the cutoff less than $1\times10^{-7}$ up to $t\cdot\tH=30$, 
$\chi_{\rm max}$ should be about 10000 as in Fig.~\ref{Ppi-t}(d).  
As shown by Fig.~\ref{Ppi-t}(c), the difference between the DMRG results for the maximum truncation
errors $1\times 10^{-7}$ and $1\times 10^{-8}$ is negligible.
Thus, it is sufficient to keep the truncation error
less than $1\times10^{-7}$ in the time-dependent DMRG simulations. 
This is computationally expensive, however, due to the rapidly increasing
bond dimensions. Alternatively, by keeping bond dimensions $\chi=800$, 
reasonable agreement can also be obtained up to $t\cdot\tH\simeq 20$ 
[see the dotted line in Fig.~\ref{Ppi-t}(c)].

\section{DMRG results}
\begin{figure}[htb]
 \includegraphics[clip,width=\textwidth]{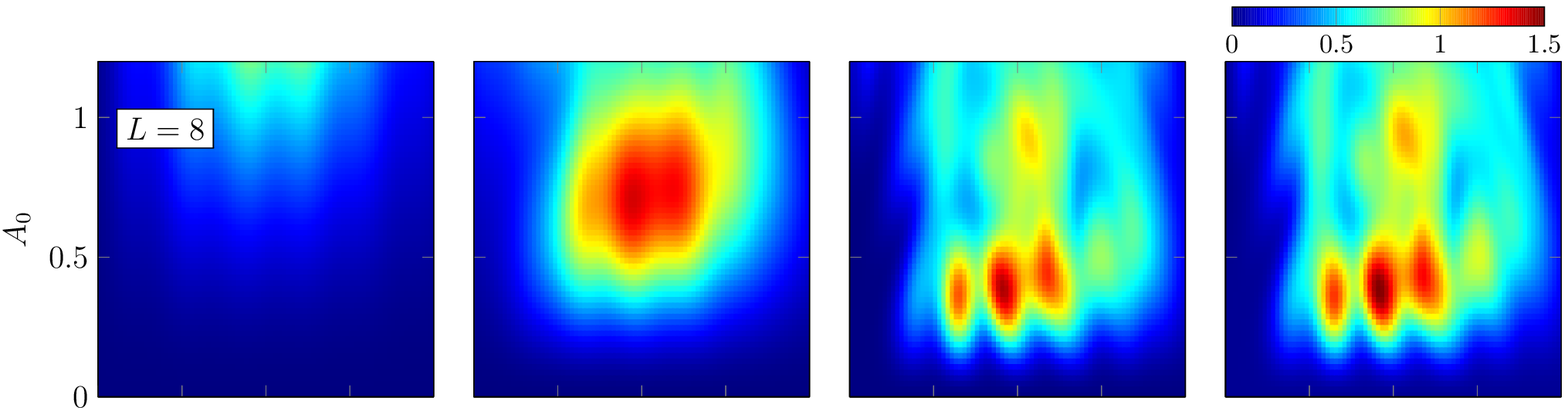}
 \includegraphics[clip,width=\textwidth]{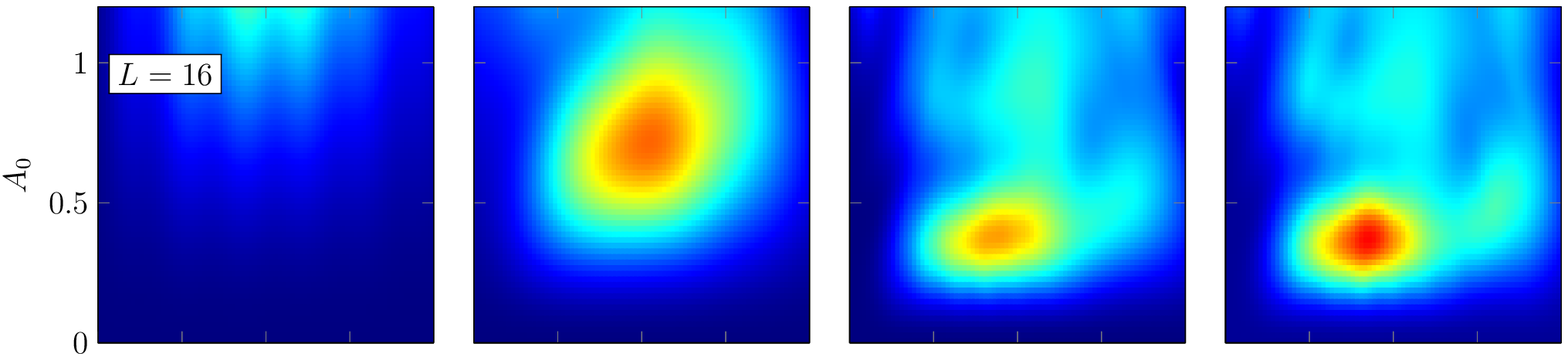}
 \includegraphics[clip,width=\textwidth]{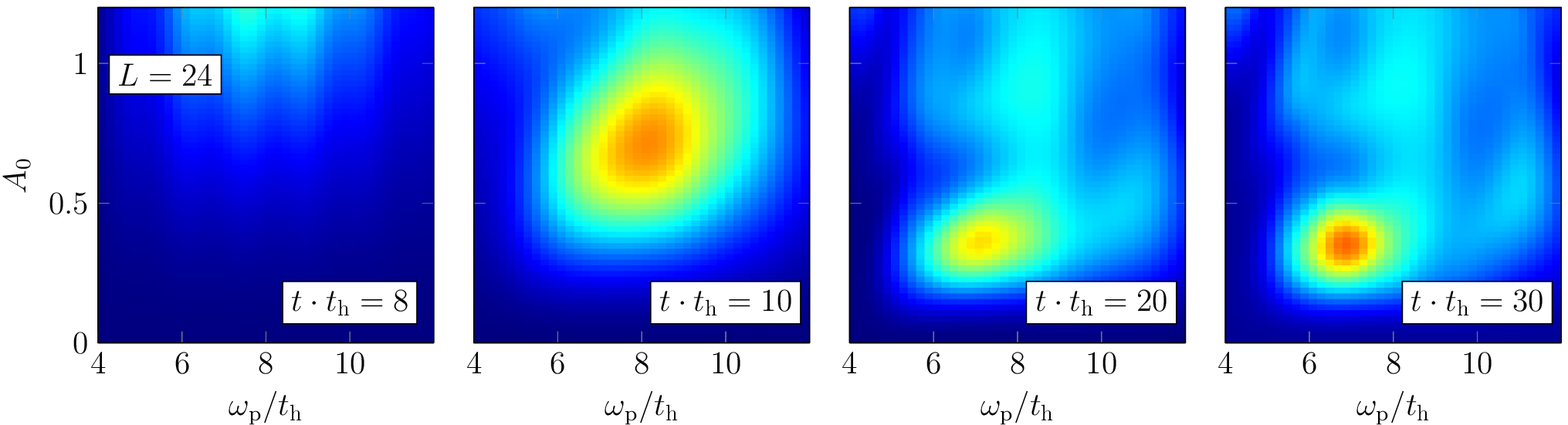}
 \caption{Contour plot of $P(q=\pi,t)$ at $t\cdot\tH=8$, $10$, $20$, 
 and $30$ for $U/\tH=8$, $\sigma_{\rm p}=2$ and $t_0\cdot\tH=10$ 
 in the $\omega_{\rm p}$-$A_0$ plane. All data obtained by DMRG with bond dimensions $\chi=800$. 
 }
 \label{CP} 
\end{figure}

We now determine the optimal parameter set for the enhancement of $P(q=\pi, t)$ 
and examine the behavior of the pair correlations for these parameters. 
Figure~\ref{CP} gives  the OBC DMRG contour plots of $P(\pi,t)$
at various times for different values of $A_0$ and $\omega_{\rm p}$. 
For $t<t_0$, the magnitude of the pair correlation functions 
is still marginal. When $t\gtrsim t_0$, a noticeable enhancement of pair correlations 
appears in wide parameter regions. 
Furthermore, the spectral intensity of $P(\pi, t)$ is concentrated 
in a single spot after pulse irradiation ($t\cdot\tH\gtrsim 20$). 
Interestingly, the height of the peak at $\omega_{\rm p}/\tH\simeq 6.8$ 
and $A_0\simeq 0.4$ increases at long times $t\cdot\tH=30$.  
Figure~\ref{CP} also demonstrates the system-size dependence of 
the pair structure factor. For $L=8$ (upper panels of Fig.~\ref{CP})
the stripe structure can be observed after the pulse irradiation 
as in the case of ED results with $L=14$~\cite{PhysRevLett.122.077002}. 
As explained in Ref.~\cite{PhysRevLett.122.077002}, the peak structure of
$P(\pi, t)$ is essentially the same as the ground-state optical spectrum $\chi_{JJ}(\omega)$
with some Lorentzian broadening $\etaL$, depending on $1/\sigma_{\rm p}$. 
If the system size is too small for some (small) $\etaL$, 
the spectrum is described by a series of peaks. The stripe structure found  in $P(\pi, t)$ 
for $L=8$ reflects this finite-size effect.
These stripes merge and construct a single peak structure 
for larger system sizes $L\geq 16$. With increasing $L$
the enhanced regime is narrowed. 
Note that the value of $\omega_{\rm p}\simeq6.8$ at the peak position  
corresponds to about the size of the Mott gap $\Delta_{\rm c}$, i.e.,
the peak position of $\chi_{JJ}(\omega)$ can be estimated to be
$\omega \sim 1.461\Delta_{\rm c}\sim 6.84$~\cite{PhysRevLett.85.3910}. 
This is in accord with the DMRG data for $L=24$, 
although Eq.~(8) of Ref.~\cite{PhysRevLett.85.3910} is only valid in 
the weak-coupling regime.

\begin{figure}[tb]
 \includegraphics[clip,width=\textwidth]{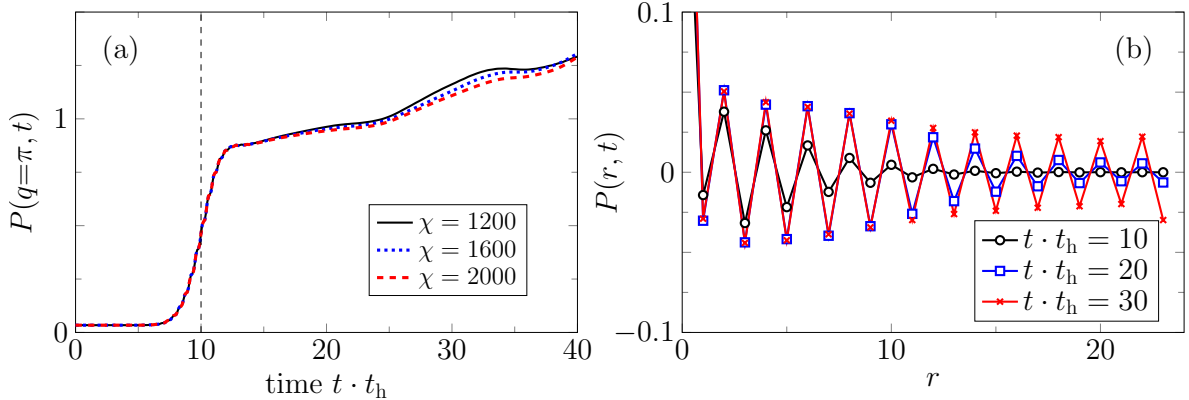}
 \caption{Time evolution of $P(q=\pi, t)$ [panel (a)] and $P(r, t)$ [panel (b)] for 
 $U/\tH=8$, $\sigma_{\rm p}=2$ and $t_0\cdot\tH=10$, where $L=24$ with OBC.  
 Parameters of the pump are $A_0=0.36$ and $\omega_{\rm p}/\tH=6.8$, 
 which corresponds to a maximum of $P(\pi, t)$ at $t\cdot \tH=30$
 in Fig.~\ref{CP}.
 }
 \label{Ppit-Prt-L24}
\end{figure}

Let us analyze the pair correlations at the peak position
of the contour plot for $L=24$ and $t\cdot\tH=30$, i.e., 
$A_0=0.36$ and $\omega_{\rm p}/\tH=6.8$. Obviously, after the pulse irradiation,
$P(\pi, t)$ 
keeps increasing gradually for $t\cdot\tH\gtrsim20$, 
as shown in Fig.~\ref{Ppit-Prt-L24}(a), 
while $P(\pi, t)$ by ED with PBC saturates at some constant value~\cite{PhysRevLett.122.077002}. 
Note that the results do not depend strongly on bond dimensions 
for large times. 
Figure~\ref{Ppit-Prt-L24}(b) displays the real-space pair correlation function.
$P(r, t)$ for $t\cdot\tH=30$ is clearly enhanced for large distances $r\gtrsim L/2$
comparing with those for $t\cdot\tH=20$.

Of course, $L=24$ is still too small to examine the behavior of
correlation functions and boundary effects are showing up in $P(r, t)$ 
for large distances. 
Moreover, the definition of the pair correlation with OBC, Eq.~\eqref{Eq-Prt}, 
is not equivalent to the usual ones with PBC
in Ref.~\cite{PhysRevLett.122.077002}. 
These data, however, might imply the possibility of a (quasi-)long-range 
order of $\eta$-pairs, 
i.e., the power-law decay of pair correlations for large times. 
Further studies for larger system sizes are desirable.

\section{Summary}
In conclusion, we demonstrated $\eta$-pairing in the one-dimensional Hubbard model 
at half filling by means of unbiased density-matrix renormalization group simulations.  
For small system sizes, the time evolution of the corresponding pair correlation
functions can be computed with high accuracy---i.e., the maximum truncation 
error is always less than $1\times 10^{-8}$---up to times $t\cdot\tH=80$ (30) 
for $L=12$ ($16$) with open boundary conditions. 
Although the numerical accuracy will get worse as $L$ increases, 
the resonant parameter set can be determined up to $L=24$. 
For these pump parameters the pair structure factor $P(q=\pi,t)$ is enhanced 
and magnitude of the pair correlation functions increases for long distances 
with time after pulse irradiation. This can be taken as strong indication for $\eta$-pairing 
and off-diagonal (quasi) long-range order. Since boundary effects still show up 
in the  pair correlation simulation data, it is highly desirable to prove 
quasi-long-range order for larger system sizes or, even better, 
directly in the thermodynamic limit by using, e.g.,  
the infinite time-evolving block decimation 
approach~\cite{PhysRevLett.98.070201}.

\section*{Acknowledgments}
We thank S. Miyakoshi and T. Shirakawa for useful discussions. 
T.K. was supported by the JSPS Overseas Research Fellowship, 
F.L. by Deutsche For\-schungsgemeinschaft(Germany) through Project 
No. FE 398/8-1, and S.Y. by Grants-in-Aid for Scientific Research from
JSPS (Project No.: JP18H01183) of Japan.
The DMRG simulations were performed using the ITensor library~\cite{ITensor}.


\end{document}